\documentclass[letterpaper, 10 pt, conference]{ieeeconf}    \IEEEoverridecommandlockouts                                                          \let\labelindent\relax \usepackage{enumitem}  
\usepackage{mathtools,amsthm}

\usepackage{graphics} 
\usepackage{epsfig}
\usepackage[linesnumbered,boxed,ruled,commentsnumbered]{algorithm2e}
\usepackage{times} 
\usepackage{amsmath,amssymb,amsfonts}
\usepackage{mathrsfs}
\usepackage{siunitx}
\usepackage{algorithmic}
\usepackage{graphicx}
\usepackage{textcomp}
\usepackage{cuted}
\usepackage{lipsum}
\usepackage{varioref}
\usepackage{import}
\usepackage{framed,xcolor}
\usepackage{color}
\usepackage{comment}
\usepackage{multirow}
\usepackage{epstopdf}   
\usepackage{psfrag}
\usepackage{breqn}
\usepackage{float}
\usepackage{mathtools}
\usepackage{booktabs}
\usepackage{soul}
\usepackage{amsbsy}
\usepackage[bottom]{footmisc}
\usepackage{lineno}
\usepackage{cleveref}
\usepackage{microtype}
\usepackage{comment}
\usepackage{geometry}

\definecolor{arash}{rgb}{0.8,0.8,1}
\definecolor{seb}{rgb}{0.8,1,0.8}
\definecolor{seb2}{rgb}{0.5,.5,1}
\definecolor{arash2}{rgb}{0,.5,0}
\definecolor{wenqi}{rgb}{1,.75,0.79}
\definecolor{wenqi2}{rgb}{1,.75,0.79}
\definecolor{hossein}{rgb}{0.8,1,0.8}

\graphicspath{{Figures/}}
\newcommand{\vect}[1]{\ensuremath{\boldsymbol{\mathrm{#1}}}}
\usepackage{graphicx}
\usepackage{tabularx,booktabs}
\usepackage{color}
\usepackage{multicol}
\usepackage{amsmath,amssymb}
\usepackage{amsfonts}
\usepackage{textcomp}
\usepackage{psfrag}
\usepackage{multimedia}
\usepackage{fancybox}
\usepackage{comment}
\usepackage{soul}
\makeatletter
\newcommand{\biggg}{\bBigg@{1.6}}  
\def\bigggl{\mathopen\biggg}
\makeatother

\newcounter{lastnote}

\usepackage{geometry}
\title{\LARGE \bf
MPC-based Reinforcement Learning for a Simplified Freight Mission of Autonomous Surface Vehicles}
\author{Wenqi Cai, Arash B. Kordabad, Hossein N. Esfahani, Anastasios M. Lekkas, and Sébastien Gros
\thanks{The authors are with Department of Engineering Cybernetics, Norwegian University of Science and Technology (NTNU), Trondheim, Norway. E-mail:{\tt\small\{wenqi.cai,arash.b.kordabad,hossein.n.esfah \newline ani,anastasios.lekkas,sebastien.gros\}@ntnu.no}}
}
\begin{document}
\maketitle
\thispagestyle{empty}
\pagestyle{empty}
\begin{abstract}
In this work, we propose a Model Predictive Control (MPC)-based Reinforcement Learning (RL) method for Autonomous Surface Vehicles (ASVs). The objective is to find an optimal policy that minimizes the closed-loop performance of a simplified freight mission, including collision-free path following, autonomous docking, and a skillful transition between them. We use a parametrized MPC-scheme to approximate the optimal policy, which considers path-following/docking costs and states (position, velocity)/inputs (thruster force, angle) constraints. The Least Squares Temporal Difference (LSTD)-based Deterministic Policy Gradient (DPG) method is then applied to update the policy parameters. Our simulation results demonstrate that the proposed MPC-LSTD-based DPG method could improve the closed-loop performance during learning for the freight mission problem of ASV.
\end{abstract}
\section{introduction}
\label{sec:intro}
\par Autonomous Surface Vehicles (ASVs) are widely applied for many fields, such as freight transportation, military, search and rescue \cite{manley2008unmanned}, and therefore attract broad attention for scientific and industrial researches. Various methods have been proposed to solve the problem of operating and automating the ASV, including path following, collision avoidance, and autonomous docking \cite{gu2019antidisturbance,martinsen2020optimization}. However, designing a control strategy that could realize both collision-free path following and docking in a freight mission with time-varying disturbances is still a topic worth exploring. With the development of Machine Learning (ML), Reinforcement Learning (RL)-based control strategies are getting noticed by people, as they can exploit real data to reduce the impact of model uncertainties and disturbances.

\par Deterministic Policy Gradient (DPG), as the direct RL method, estimates the optimal policy by a parameterized function approximator, and optimizes the policy parameters directly via gradient descent steps of the performance \cite{SuttonPG}. Deep Neural Networks (DNNs) are very commonly used function approximators in RL \cite{bucsoniu2018reinforcement}. However, DNN-based RL lacks the abilities concerning the closed-loop stability analysis, state/input constraints satisfaction, and meaningful weights initialization \cite{zanon2020safe}. To address these problems, the perspective of using Model Predictive Control (MPC)-based RL has been proposed and justified in~\cite{gros2019data}, i.e. it suggests using MPC as the function approximation for the optimal policy in RL. Unlike DNNs, MPC-based policies satisfy the state/input constraints and safety requirements by construction, and its well-structured property enables the stability analysis of the system.

\par However, for computational reasons, simple models are usually preferred in the MPC-scheme. Hence, the MPC model often does not have the required structure to correctly capture the real system dynamics and stochasticity. As a result, MPC can deliver a reasonable approximation of the optimal policy, but it is usually suboptimal~\cite{camacho2013model}. Besides, choosing the model parameters that best fit the MPC model to the real system does not necessarily yield a policy that achieves the best closed-loop performance \cite{zanon2020safe}. Therefore, choosing appropriate MPC parameters to achieve the best closed-loop performance is extremely challenging. Nevertheless, according to \textit{Theorem 1} and \textit{Corollary 2} in \cite{gros2019data}, it can conclude that by adjusting not only the MPC model parameters but also the parameters in the MPC cost and constraints, the MPC scheme can, theoretically, generate the optimal closed-loop policy even if the MPC model is inaccurate. It is also shown that RL is a suitable candidate to perform that adjustment. Recent researches focused on the MPC-based RL have further developed this approach~\cite{grosreinforcement,zanon2019practical,Arash2021CCTAGrid,cai2021optimal,arash2021battery}. 

\par The contribution of this work is to provide a promising approach for a complete ASV freight mission problem. The problem is challenging since it needs to solve the obstacle avoidance, path following, and autonomous docking simultaneously, in a stochastic environment. We elaborate the proposed MPC-based RL method in the ASV problem framework, as well as formulate an algorithm for the MPC-LSTD-based DPG method.
\section{ASV model}
\label{sec:model}
\begin{figure}[htbp!]
\centering
\includegraphics[width=0.3\textwidth]{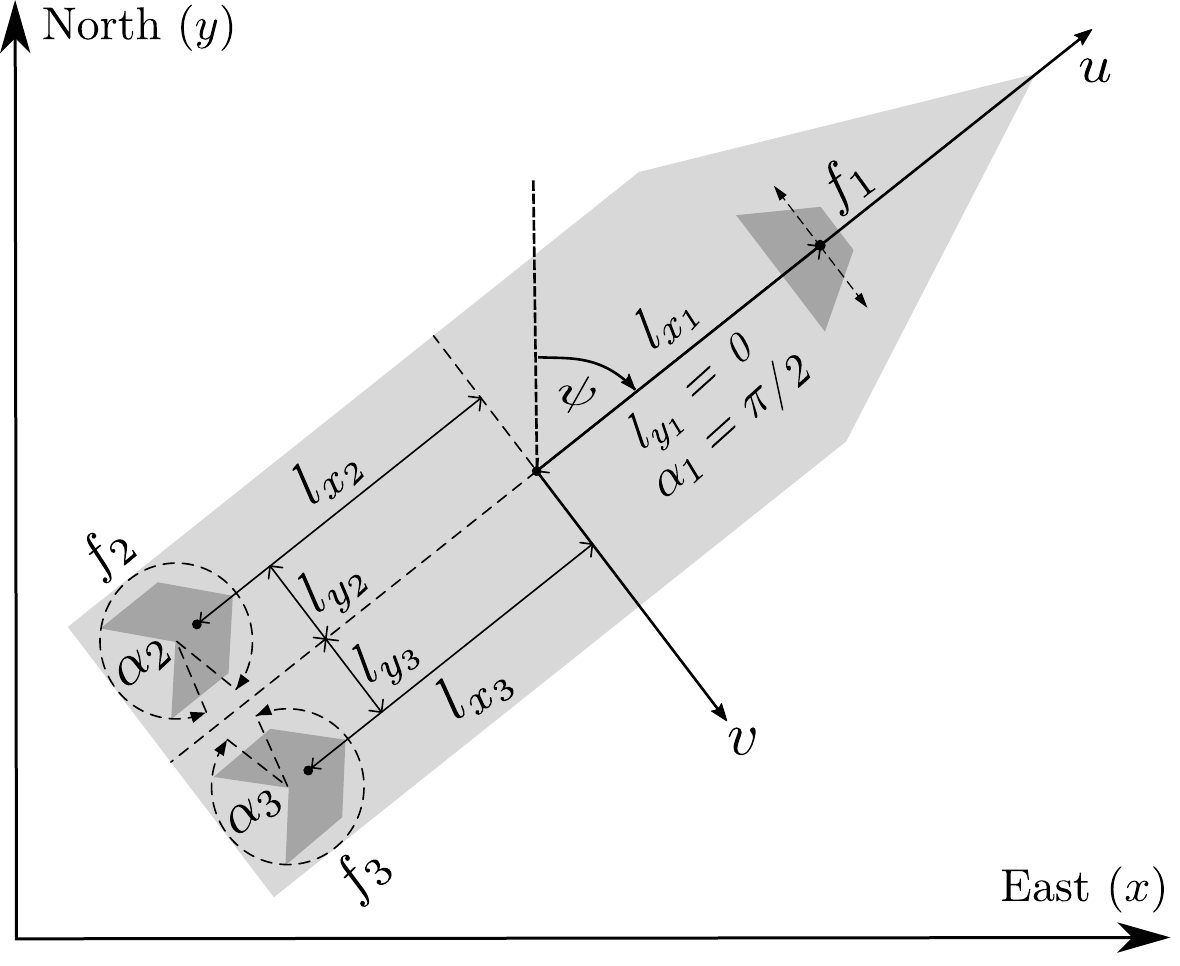}
\caption{The 3-DOF ASV model in the NED frame.}
\label{figdyn}
\end{figure}

The 3-Degree of Freedom (3-DOF) position of the vessel can be represented by a pose vector  $\boldsymbol{\eta}=[x,y,\psi]^{T}\in\mathbb{R}^3$ in the North-East-Down (NED) frame, where $x$ is the North position, $y$ is the East position, and $\psi$ is the heading angle (see Fig. \ref{figdyn}). The velocity vector $\boldsymbol{\nu}=[u,v,r]^{T}\in\mathbb{R}^3$, including the surge velocity $u$, sway velocity $v$, and yaw rate $r$, is decomposed in the body-fixed frame. 
The nonlinear dynamics can be written as follows~\cite{skjetne2004modeling}
\begin{subequations}\label{con-dyn}
\begin{gather}\label{eq:dyn1}
   \dot{\vect{\eta}}=\vect{J}(\psi)\vect{\nu} \\
    \vect M\vect{\dot{\nu}}+
     \vect D\vect{\nu}=\vect\tau+\vect\tau_a,
\end{gather}
\end{subequations}
where $\vect{J}(\psi)\in\mathbb{R}^{3\times 3}$ is the rotation matrix, $\vect{M}\in\mathbb{R}^{3\times 3}$ is the mass matrix, and $\vect{D}\in\mathbb{R}^{3\times 3}$ is the damping matrix (see \cite{martinsen2019autonomous} for their specific physical meanings and values). Vector $\vect\tau\in\mathbb{R}^3$ presents the control forces and moment empowered by the thrusters. Vector $\vect\tau_a\in\mathbb{R}^3$ is the additional forces rendered from disturbances, e.g., wind, ocean wave and etc. The thrust configuration is illustrated in Fig. \ref{figdyn}. The vector $\vect\tau$ could be specifically written as $\vect\tau=\vect T\left(\vect\alpha\right)\vect f$, where $\vect f=[f_1,f_2,f_3]^\top\in\mathbb{R}^3$ is the thruster forces vector as we consider one tunnel thruster $f_1$ and two azimuth thrusters $f_2,f_3$. They are subjected to the bounds
\begin{gather}
{f_p}_{\mathrm{min}} \leq f_p\leq {f_p}_{\mathrm{max}},\quad p=1,2,3. \label{eg:f bound}
\end{gather}
Matrix $\vect T\left(\vect\alpha\right)\in\mathbb{R}^{3\times 3}$ presents the thruster configuration, written as
\begin{gather}
\small
\label{eq:T(alpha)}
   \vect T\left(\vect\alpha\right)= \begin{bmatrix}
    0 & \cos\left(\alpha_2\right) & \cos\left(\alpha_3\right) \\ 
    1 &\sin\left(\alpha_2\right) & \sin\left(\alpha_3\right)\\
    {l_x}_1& T_{32} & T_{33}
    \end{bmatrix},
\end{gather}
where elements $T_{32}={l_x}_2\sin\left(\alpha_2\right)-{l_y}_2\cos\left(\alpha_2\right)$, and $T_{33}={l_x}_3\sin\left(\alpha_3\right)-{l_y}_3\cos\left(\alpha_3\right)$. Constants ${l_x}_{i}$ and ${l_y}_{i}$ with $i=1,2,3$ are the distances between each thruster and the cross line of the ship's center. Term $\vect \alpha=[\alpha_1,\alpha_2,\alpha_3]^\top\in\mathbb{R}^3$ is the corresponding orientation vector. The angle $\alpha_1$ is fixed ($\pi/2$), while $\alpha_2$ and $\alpha_3$, associated to the two azimuth thrusters, are restricted in the range
\begin{align}
\label{eg:angle bound}
\left|\alpha_2+\pi/2\right|\leq {\alpha}_{\mathrm{max}}, \quad \left|\alpha_3-\pi/2\right|\leq {\alpha}_{\mathrm{max}}.
\end{align}
A maximum angle of ${\alpha}_{\mathrm{max}}$ with a forbidden sector is considered in this work to avoid thrusters $2$ and $3$ directly work against each other, as shown in Fig. \ref{figdyn}. With a sampling time of $dt$, we discretize the ship system \eqref{con-dyn} as
\begin{align}
    \vect s_{k+1}=F\left(\vect s_{k},
    \vect a_{k},\vect\tau_a\right),
    \label{eq:df}
\end{align}
where $\vect s_k=\left[\vect\eta_k^\top,\vect\nu_k^\top\right]^\top$ and $\vect a_k=\left[\vect f^\top_k,\vect\alpha^\top_k\right]^\top$ are system state and input vectors, respectively. Subscript $k$ denotes the physical time and $F(\cdot)$ is the discretized real system.
\section{Simplified Freight Mission}
\label{sec:Mission}
In this work, we consider a simplified freight mission problem: the ASV starts from an origin $\vect A$ to the end $\vect B$, which is supposed to follow a designed collision-free course and finally dock at the wharf autonomously. Note that the transition from path following to docking is a notable point of this problem.
\subsection{Collision-Free Path Following}
\label{sec:path following}
\par Given a reference path $P_{\mathrm{ref}}$. At time instance $k$, $\vect P^{\mathrm{ref}}_k=[x^{\mathrm{ref}}_k, y^{\mathrm{ref}}_k]^\top$. Then path following could be thought as minimizing the error $l\left( \vect\eta_k \right)$
\begin{align}
    l\left( \vect\eta_k \right) ={\left\| {\vect\eta}^{p}_k -\vect P^{\mathrm{ref}}_k \right\|}_2^2= (x_k - x^{\mathrm{ref}}_k)^2 + (y_k - y^{\mathrm{ref}}_k)^2,
\end{align}
where ${\vect\eta}^{p}_k=[x_k,y_k]^\top$ contains the first two elements of $\vect\eta_k$. Besides, we assume obstacles of round shape. To avoid these obstacles, the following term $g_n\left(\vect\eta_k\right)$, representing the position of the ship relative to the $n^{\mathrm{th}}$ obstacle, should satisfy
\begin{gather}
\left(x_k-o_{x,n}\right)^2+\left(y_k-o_{y,n}\right)^2\geq \left(r_{n}+r_o\right)^2, \label{obs}
\end{gather}
i.e., 
\begin{gather}
\underbrace {1 - \left( {{{\left( {{x_k} - {o_{x,n}}} \right)}^2} + {{\left( {{y_k} - {o_{y,n}}} \right)}^2}} \right)\bigggl/{{\left( {{r_n} + {r_o}} \right)}^2}}_{g_n\left( {{\vect \eta _k}} \right)} \le 0,
\label{eq:obs_penalty}
\end{gather}
where $(o_{x,n},o_{y,n})$ and $r_n$ are the center and radius of the $n^{\mathrm{th}}$ circular obstacle ($n=1,\ldots,N_o$), respectively. Constant $r_o$ is the radius of the vessel and $N_o$ is the number of obstacles.
\subsection{Autonomous Docking}
\label{sec:docking}
Docking refers to stopping the vessel exactly at the endpoint $\vect B$ as well as avoiding collisions between any part of the vessel and the quay \cite{martinsen2020optimization}. The ``accurate stop" requires not only an accurate docking position but also zero-valued velocities and thruster forces at the final time, i.e., we ought to minimize
\begin{align}
    h\left(\vect\eta_k,\vect\nu_k,\vect f_k\right)=\left\|\vect\eta_k - \vect\eta_d\right\|^2_2 + \left\|\vect\nu_k\right\|^2_2 +\left\|\vect f_k\right\|^2_2,
\end{align}
where $\vect\eta_d=(x_d,y_d,\psi_d)$ is the desired docking position. Successfully docking requires $h\left(\vect\eta_K,\vect\nu_K,\vect f_K\right)\approx 0$, where subscript $K$ denotes the terminal time step of the freight mission. As for ``collision avoidance", we define a safety operation region $\mathbb{S}$ as the spatial constraints for the vessel. The operation region is chosen as the largest convex region that encompasses the docking point but not intersecting with the land. Thus, as long as the vessel is within the region $\mathbb{S}$, no collision will occur during docking, i.e. the following condition should hold
\begin{align}
    {\vect\eta}^{p}_k\in \mathbb{S},\quad\mathbb{S}=\{\vect x|\vect A\vect x<\vect b\},
\end{align}
where ${\vect\eta}^{p}_k=[x_k,y_k]^\top$ describes the position of the vessel. The matrix $\vect A$ and the vector $\vect b$ are determined by the shape of the quay and together define the convex region $\mathbb{S}$.
\subsection{Objective Function}
\label{sec:objective}
In the context of RL, we seek a control policy $\vect\pi$ that minimizes the following closed-loop performance $J$
\begin{align}
\label{eq:J}
J(\vect\pi)=\mathbb E_{\vect\pi}\Bigg[\sum_{k=0}^{K} \gamma^k L\left( \vect s_k,\vect a_k\right)\Bigg|\vect a_k=\vect \pi(\vect s_k)\Bigg],
\end{align}
where $\gamma\in(0,1]$ is the discount factor. Expectation $\mathbb E_{\vect\pi}$ is taken over the distribution of the Markov chain in the closed-loop under policy $\vect\pi$. The RL-stage cost $L(\vect s_k,\vect a_k)$, in this problem, is defined as a piecewise function:
\begin{align}
\label{eq:rl cost}
L = \left\{ {\begin{array}{*{20}{l}}
{l\left( {{\vect\eta _k}} \right)+O\left( {{\vect\eta _k}} \right)+\xi\left( \vect\alpha_k  \right)}&{{{\left\| {\vect\eta_k  - {\vect\eta _d}} \right\|}^2_2} > d}\\
{h\left( {{\vect\eta _k},{\vect\nu _k},{\vect f_k}} \right)+\Gamma\left( {{\vect\eta _k}} \right)+\xi\left( \vect\alpha_k  \right)}&{{{\left\| {\vect\eta_k  - {\vect\eta _d}} \right\|}^2_2} \leq d},
\end{array}} \right.
\end{align}
where $O\left( {{\vect\eta _k}} \right)$ is the obstacle penalty for path following
\begin{align}
\label{eq:safe}
   O\left( {{\vect\eta _k}} \right)= \sum_{n=1}^{N_o}c_{n}\cdot\max(0,g_n\left( {{\vect\eta _k}} \right)+d_s),
\end{align}
where $c_{n}>0$ is the penalty weight, constant $d_s>0$ is the desired safe distance between vessel and obstacles. Therefore, once the ship breaks the safe distance, i.e. $g_n\left( {{\vect\eta _k}} \right)+d_s>0$, a positive penalty will be introduced to the objective function. Function $\Gamma\left( {{\vect\eta _k}} \right)$ is the collision penalty for docking
\begin{align}
\label{eq:docking_penalty}
   \Gamma\left( {{\vect\eta _k}} \right)=\kappa\cdot (1-\vect1_{\mathbb{S}}(\vect\eta _k^p)),
\end{align}
where $\kappa>0$ is the penalty weight and $\vect 1_{\mathbb{S}}(\cdot)$ is the indicator function. When the ship is out of the safe region, i.e. ${\vect\eta}^{p}_k\notin \mathbb{S}$, a positive penalty will be imposed in the objective function. Function $\xi\left( \vect\alpha_k  \right)$ is the singular configuration penalty, aiming to avoid the thruster configuration matrix $\vect T\left( \vect\alpha_k  \right)$ in \eqref{eq:T(alpha)} being singular \cite{johansen2004constrained}
\begin{align}
\small
   \xi\left( \vect\alpha_k  \right)=\frac{\rho }{{\varepsilon  + \det \left( {\vect T\left( \vect\alpha_k  \right)\vect W^{-1}{\vect T^\top}\left( \vect\alpha_k  \right)} \right)}},
\end{align}
where ``$\det$" stands for the determinant of the matrix. Constant $\varepsilon>0$ is a small number to avoid division by zero, $\rho>0$ is the weighting of maneuverability, and $\vect W$ is a diagonal weighting matrix. Constant $d>0$ is designed to substitute the stage cost from path following to docking at ${{\left\| {\vect\eta_k  - {\vect\eta _d}} \right\|}^2_2}=d$, which means that our target transits from path-following to docking when the ship approaches the destination.
\section{MPC-Based Reinforcement Learning}
\label{sec:MPC+PG}
The core idea of our proposed approach is to use a parameterized MPC-scheme as the policy approximation function, and apply the LSTD-based DPG method to update the parameters so as to improve the closed-loop performance.
\subsection{MPC-Based Policy Approximation}
\label{sec:MPC}
Consider the following MPC-scheme parameterized with $\vect\theta$
\begin{subequations}
\label{eq:mpc}
\begin{align}
\nonumber \min_{\small{\hat{\vect\eta}, \hat{\vect\nu},\hat{\vect f},\hat{\vect\alpha},\vect\sigma}}&\, \frac{\theta_d}{{{{\left\| {{\hat{\vect\eta} _N} - {\vect\eta _d}} \right\|}^2_2} + \delta }}\cdot \Big( h_{\vect\theta}\left(\hat{\vect\eta}_N,\hat{\vect\nu}_N\right)  +\Gamma_{\theta}\left( {\hat{\vect\eta} _N} \right)\Big)+\\
\label{eq:mpc_cost}
& {\vect\omega}_{f}^\top {\vect\sigma}_{N} +\sum_{i=0}^{N-1} \gamma^i\Big(l_{\vect\theta}\left(\hat{\vect\eta}_i\right) + \xi\left(\hat{\vect\alpha}_i \right)+ {\vect\omega}^\top {\vect\sigma}_{i}\Big) \\
\mathrm{s.t.}\quad &\forall i=0,\ldots,N-1, \,\, n=1,\ldots,N_o \nonumber\\
\label{eq:mpc_1}
&\left[\hat{\vect\eta}{^\top}_{i+1},\hat{\vect\nu}{^\top}_{i+1}\right]{^\top}=F_{\vect\theta}(\hat{\vect\eta}_{i},\hat{\vect\nu}_{i}, \hat{\vect f}_{i},\hat{\vect\alpha}_{i},\vect\theta_a)\\
\label{eq:mpc_2}
&{f_p}_{\mathrm{min}} \leq \hat f_{p,i}\leq {f_p}_{\mathrm{max}},\quad p=1,2,3\\
\label{eq:mpc_3}
&\left|\hat\alpha_{2,i}+\pi/2\right|\leq {\alpha}_{\mathrm{max}}, \,\,\, \left|\hat\alpha_{3,i}-\pi/2\right|\leq {\alpha}_{\mathrm{max}},\\
\label{eq:mpc_4}
&g_n\left(\hat{\vect\eta}_i\right)+\theta_{g}\leq \sigma_{n,i},\,\,\, g_n\left(\hat{\vect\eta}_N\right)+\theta_{g}\leq \sigma_{n,N},\\
\label{eq:mpc_6}
&\vect\sigma_{i}\geq 0 , \,\,\, \vect\sigma_{N}\geq 0, \\
&\hat{\vect\eta}_0=\vect\eta_k,\,\,\, \hat{\vect\nu}_0=\vect\nu_k, 
\end{align}
\end{subequations}
where $N$ is the prediction horizon. Arguments $\hat{\vect\eta}=\{\hat{\vect\eta}_{0},\dots,\hat{\vect\eta}_{N}\}$, $\hat{\vect\nu}=\{\hat{\vect\nu}_{0},\dots,\hat{\vect\nu}_{N}\}$, $\hat{\vect f}=\{\hat{\vect f}_{0},\dots,\hat{\vect f}_{N-1}\}$, $\hat{\vect\alpha}=\{\hat{\vect\alpha}_{0},\dots,\hat{\vect\alpha}_{N-1}\}$, and ${\vect\sigma}=\{\vect\sigma_{0},\dots ,\vect\sigma_{N}\}$ are the primal decision variables. The term $\frac{\theta_d}{{{{\left\| {{\hat{\vect\eta} _N} - {\vect\eta _d}} \right\|}^2_2} + \delta }}\cdot \left(h_{\vect\theta}\left(\cdot\right)+\Gamma_{\vect\theta}\left(\cdot\right)\right)$ introduces a gradually increasing terminal cost as the ship approaches the endpoint, where $\delta>0$ is a small constant to avoid division by zero. The weighting parameter $\theta_d$, designed to balance the priority of path following and docking, is tuned by RL. Note that $\theta_d$ is chosen to minimize the closed-loop performance that considering both path following and docking, although it may be suboptimal for either single problem. Parameter $\theta_g$ is the tightening variable used to adjust the strength of the collision avoidance constraints. If the value of $\theta_g$ (positive) is larger, it means that the constraints are tighter and the ship is supposed to be farther away from the obstacles. It is important to use RL to pick an appropriate $\theta_g$, since when $\theta_g$ is too large, although we ensure that the ship safely avoids obstacles, the path following error is increased. Conversely, a smaller $\theta_g$ reduces the following error, but we may gain more penalty when the vessel breaks the safe distance, as described in \eqref{eq:safe}. Note that the obstacle penalties are considered directly as constraints \eqref{eq:mpc_4} in the MPC rather than as penalties in the MPC cost, because \eqref{eq:obs_penalty} is a conservative model of the obstacle penalty \eqref{eq:safe}. Variables $\vect\sigma_{i}$ $\big(\vect\sigma_{i}=\{\sigma_{1,i},\dots,\sigma_{N_0,i}\}\big)$ and $\vect\sigma_{N}$ $\big(\vect\sigma_{N}=\{\sigma_{1,N},\dots,\sigma_{N_0,N}\}\big)$ are slacks for the relaxation of the state constraints, weighted by the positive vectors $\vect\omega$ and $\vect\omega_f$. The relaxation prevents the infeasibility of the MPC in the presence of some hard constraints.

\par The parameterized stage cost $l_{\vect\theta}\left(\cdot\right)$, terminal cost $h_{\vect\theta}\left(\cdot\right)$, and docking collision penalty $\Gamma_{\theta}\left(\cdot\right)$ in the MPC cost \eqref{eq:mpc_cost} are designed as follows
\begin{subequations}\label{eq:approx}
\begin{align}
 l_{\vect\theta} &={\left\| {\hat{\vect\eta}}^p_i -\vect P^{\mathrm{ref}}_i \right\|}_{\vect\Theta_{l}}^2\\
h_{\vect\theta}&=\left\|\hat{\vect\eta}_N - \vect\eta_d\right\|_{\vect\Theta_{\eta}}^2 + \left\|\hat{\vect\nu}_N\right\|_{\vect\Theta_{\nu}}^2 \\
\Gamma_{\theta}&=\theta_{\kappa}\cdot(1-\vect1_{\mathbb{S}}({{\hat{\vect\eta}}_N^p})),
\end{align}
\end{subequations}
where $\vect\Theta_l,\vect\Theta_{\eta}, \vect\Theta_{\nu}\in\mathbb{R}^{3\times 3}$ are the weighing matrices that are symmetric semi-positive definite. They are expressed as $\vect\Theta_l=(\mathrm{diag}(\vect\theta_l))^2$, $\vect\Theta_\eta=(\mathrm{diag}(\vect\theta_\eta))^2$, $\vect\Theta_\nu=(\mathrm{diag}(\vect\theta_\nu))^2$. Operator ``$\mathrm{diag}$" assigns the vector elements onto the diagonal elements of a square matrix. Parameter $\theta_\kappa$ is treated as a degree of freedom for the docking collision penalty. The real model is \eqref{eq:df} and we assume the disturbance $\vect\tau_a$ follows a Gaussian distribution. To address the disturbance without using a complex stochastic model in the MPC scheme, one measure is to use a parameter vector $\vect\theta_a\in\mathbb{R}^3$ to parameterize the model as $F_{\vect\theta}(\hat{\vect s}_{
i}, \hat{\vect a}_{i},\vect\theta_a)$. As detailed in \cite{gros2019data}, the full adaptation of the parametrized MPC scheme (model, costs, constraints) can compensate for that unmodelled disturbance. Overall, the adjustable parameters vector $\vect\theta$ is consisted as
\begin{align}
    \vect\theta=\{\vect\theta_l,\vect\theta_{\eta},\vect\theta_{\nu},\vect\theta_a,\theta_\kappa,\theta_d,\theta_g\}.
\end{align}
And $\vect\theta$ will be adjusted by RL according to the principle of ``improving the closed-loop performance". Note that: 1. the span of the RL ($K\approx550$) is much longer than the horizon of the MPC ($N=60$); 2. the RL cost \eqref{eq:rl cost} is a ``switching" function, while the MPC cost \eqref{eq:mpc_cost} contains simultaneously the path following and docking cost to avoid the mixed-integer treatment of the problem; 3. the MPC model does not perfectly match the real system. For the above reasons, having different cost functions in the MPC scheme and RL is rational \cite{gros2019data}. Therefore, in order to improve the closed-loop performance of the MPC scheme as assessed by the RL cost, it can be beneficial to parameterize the MPC cost functions, model, and constraints. RL then adjusts these parameters according to the principle of ``improving the closed-loop performance". From \textit{Theorem 1} and \textit{Corollary 2} in \cite{gros2019data}, we know that, theoretically, under some assumptions, if the parametrization is rich enough, the MPC scheme is capable of capturing the optimal policy $\pi^\star$ in presence of model uncertainties and disturbances.

\par Importantly, the deterministic policy  $\vect \pi_{\vect \theta}(\vect s)$ can be obtained as
 \begin{align}
    \vect \pi_{\vect \theta}(\vect s)=\vect u_{0}^{\star}(\vect s,\vect \theta),
\end{align}
where $\vect u_{0}^{\star}(\vect s,\vect \theta)$ is the first element of ${\vect u}^\star$, which is the input solution of the MPC scheme \eqref{eq:mpc}.
\subsection{LSTD-Based DPG Method}
\label{sec:PG}
The DPG method optimizes the policy parameters $\vect\theta$ directly via gradient descent steps on the performance function $J$, defined in \eqref{eq:J}. The update rule is as follows
\begin{align}
\label{eq:update theta}
    \vect\theta \leftarrow \vect\theta-\alpha  \nabla _{\vect\theta}J(\vect\pi _{\vect\theta}),
\end{align}
where $\alpha>0$ is the step size. Applying the DPG method developed by \cite{silver2014deterministic}, the gradient of $J$ with respect to parameters $\vect\theta$ is obtained as
\begin{align}\label{eq:dj}
    \nabla _{\vect\theta}J(\vect\pi _{\vect\theta}) = \mathbb E\left[{\nabla _{\vect\theta} }{\vect\pi _{\vect\theta} }(\vect s){\nabla _{\vect a}}{Q_{{\vect\pi _{\vect\theta} }}}(\vect s,\vect a)|_{\vect a=\vect \pi _{\vect\theta}}\right],
\end{align}
where $Q_{\vect{\pi}_{\vect{\theta}}}$ and its inner function $V_{\vect{\pi}_{\vect{\theta}}}$ are the action-value function and value function associated to the policy ${\vect{ \pi}}_{\vect{\theta}}$, respectively, defined as follows
\begin{subequations}
\begin{align}
{Q_{\vect\pi _{\vect\theta }}}\left( {\vect s,\vect a} \right) &= L\left( {\vect s,\vect a} \right) + \gamma {\mathbb E}\left[ {{V_{\vect\pi _{\vect\theta }}}\left( {{\vect s^ + }|(\vect s,\vect a)} \right)} \right] \label{eq:Q}\\
{V_{\vect\pi _{\vect\theta }}}(\vect s) &= {Q_{\vect\pi _{\vect\theta }}}\left( \vect s,{\vect\pi _{\vect\theta }\left( \vect s \right)} \right), \label{eq:V}
\end{align}
\end{subequations}
where $\vect s^+$ is the subsequent state of the state-input pair ($\vect s, \vect a$). The calculations of ${\nabla _{\vect\theta} }{\vect\pi _{\vect\theta} }(\vect s)$ and ${\nabla _{\vect a}}{Q_{{\vect\pi _{\vect\theta} }}}(\vect s,\vect a)$ in (\ref{eq:dj}) are discussed in the following.
\subsubsection{${\nabla _{\vect\theta} }{\vect\pi _{\vect\theta} }(\vect s)$}
The primal-dual Karush Kuhn Tucker (KKT) conditions underlying the MPC scheme \eqref{eq:mpc} is written as
\begin{align}
    \vect R = {\left[ {\begin{array}{*{20}{c}}
{{\nabla _{\vect \zeta}}{\mathcal L_{\vect\theta} }}&{{\vect G_{\vect\theta} }}&{\mathrm{diag}\left(\vect\mu\right) \vect H_{\vect\theta} }
\end{array}} \right]^\top},
\end{align}
where $\vect\zeta=\{\hat{\vect\eta}, \hat{\vect\nu},\hat{\vect f},\hat{\vect\alpha},\vect\sigma\}$ is the primal decision variable of the MPC \eqref{eq:mpc}. Term $\mathcal{L}_{\vect \theta}$ is the associated Lagrange function, written as
\begin{align}
\mathcal{L}_{\vect \theta}(\vect y) = \Omega_{\vect \theta} + \vect\lambda^\top \vect G_{\vect\theta}  + \vect\mu^\top \vect H_{\vect \theta},
\end{align}
where $\Omega_{\vect\theta}$ is the MPC cost \eqref{eq:mpc_cost}, $\vect G_{\vect\theta}$ gathers the equality constraints and $\vect H_{\vect\theta}$ collects the inequality constraints of the MPC \eqref{eq:mpc}. Vectors $\vect\lambda,\vect\mu$ are the associated dual variables. Argument ${\vect y}$ reads as ${\vect y} =\{\vect\zeta,\vect\lambda,\vect\mu\}$ and $ {\vect y}^{\star}$ refers to the solution of the MPC \eqref{eq:mpc}. Consequently, the policy sensitivity ${\nabla _{\vect \theta} }{\vect \pi _{\vect \theta} }$ required in \eqref{eq:dj} can then be obtained as follows~(\cite{gros2019data})
\begin{align}
\label{eq:sensetivity}
{\nabla _{\vect \theta} }{\vect \pi _{\vect \theta} }\left(\vect  s \right) =  - {\nabla _{\vect\theta} }{\vect R }\left( {\vect y^\star},\vect s,\vect\theta\right){\nabla _{\vect y}}{\vect R }{\left( {\vect y^\star},\vect s,\vect\theta \right)^{ - 1}}\frac{\partial {\vect y}}{\partial {\vect u_0}},
\end{align}
where $\vect u_0$ is the first element of the input, expressed as
\begin{align}
    \vect u_0=\left[\vect {\hat f}^\top_0,\hat{\vect\alpha}^\top_0\right]^\top.
\end{align}
\subsubsection{${\nabla _{\vect a}}{Q_{{\vect\pi _{\vect\theta} }}}(\vect s,\vect a)$}
Under some conditions~\cite{silver2014deterministic}, the action-value function $Q_{{\vect\pi _{\vect\theta} }}$ can be replaced by an approximator ${Q_{\vect w}}$, i.e. $Q_{\vect w}\approx Q_{\vect \pi_{\vect \theta}}$, without affecting the policy gradient. Such an approximation is labelled \textit{compatible} and can, e.g., take the form
\begin{align}
\label{eq:Q_w}
{Q_{\vect w}}\left( {\vect s,\vect a} \right) = \underbrace {{{\left( {\vect a - {\vect\pi _{\vect\theta} }\left( \vect s \right)} \right)}^{\top}}{\nabla _{\vect\theta} }{\vect\pi _{\vect\theta} }{{\left( \vect s \right)}^{\top}}}_{{\vect\Psi ^{\top}}\left( {\vect s,\vect a} \right)}\vect w + { V_{\vect v}}\left( \vect s \right),
\end{align}
where $\vect\Psi(\vect s,\vect a)$ is the state-action feature vector, $\vect w$ is the parameters vector estimating the action-value function $Q_{{\vect{ \pi}}_{\vect{\theta}}}$ and $V_{\vect v}\approx V_{\vect \pi_{\vect \theta}}$ is the parameterized baseline function approximating the value function, it can take a linear form
\begin{align}
\label{eq:V_v}
    {V_{\vect v}} \left(\vect s\right ) =\vect \Phi\left(\vect s \right)^\top {\vect v},
\end{align}
where $\vect \Phi(\vect s)$, the state feature vector, is designed to constitute all monomials of the state with degrees less than or equal to $2$. And $\vect v$ is the corresponding parameters vector. Now we get
\begin{align}
    {\nabla _{\vect a}}{Q_{{\vect\pi _{\vect\theta} }}}(\vect s,\vect a) \approx {\nabla _{\vect a}}{Q_{\vect w}}(\vect s,\vect a)=\nabla _{\vect\theta}\vect \pi _{\vect\theta}\left(\vect s\right )^{\top}\vect w.
\end{align}
The parameters $\vect w$ and $\vect v$ of the action-value function approximation \eqref{eq:Q_w} are the solutions of the Least Squares (LS) problem
\begin{align}
\label{eq:error}
    \min_{\vect w, \vect v} \mathbb{E} \left[\big( Q_{\vect\pi_{\vect\theta}}(\vect s,\vect a)-Q_{\vect w} (\vect s,\vect a)\big )^2\right],
\end{align}
which, in this work, is tackled via the LSTD method (see \cite{lagoudakis2003least}). LSTD belongs to \textit{batch method}, seeking to find the best fitting value function and action-value function, and it is more sample efficient than other methods. The LSTD update rules are as follows
\begin{subequations}
\small
\label{eq:lstdq_update}
\begin{align}
\label{eq:lstdq_1}
    &{\vect v} = \mathbb E_{m} \Bigg\{{\left[ {\sum_{k=1}^{K}\left[ {\vect\Phi (\vect s_k){{\left( {\vect \Phi (\vect s_k) - \gamma \vect\Phi ({\vect s_{k+1} })} \right)}^\top}} \right]} \right]^{ - 1}}\nonumber\\&\qquad\quad\qquad\qquad\qquad\qquad\qquad{\sum_{k=1}^{K}}\Big[ {\vect\Phi (\vect s_k)L(\vect s_k,\vect a_k)} \Big]\Bigg\},\\
\label{eq:lstdq_2}
    &{\vect w} =\mathbb E_{m} \Bigg\{ {\left[ {{\sum_{k=1}^{K}}\Big[ \vect\Psi(\vect s_k,\vect a_k)\vect\Psi(\vect s_k,\vect a_k)^{\top} \Big]} \right]^{ - 1}}\nonumber\\
    &{\sum_{k=1}^{K}}\Big[ {\left( {L(\vect s_k,\vect a_k) + \gamma {V_{\vect v}}\left( {{\vect s_ {k+1}}} \right) - {V_{\vect v}}\left( \vect s_k \right)} \right)\vect\Psi(\vect s_k,\vect a_k)} \Big]\Bigg\},
\end{align}
\end{subequations}
where the summation is taken over the whole episode, which terminates at $K$ when the ship reaches the destination (i.e. ${{{\left\| {\vect\eta_K  - {\vect\eta _d}} \right\|}^2_2} \leq d_{\mathrm{error}}}$). The values will be then averaged by taking expectation ($\mathbb{E}_m$) over $m$ episodes.

Finally, equation \eqref{eq:update theta} can be rewritten as a compatible DPG
\begin{align}
\label{eq:new_theta}
    {\vect \theta} \leftarrow {\vect\theta} - \alpha \mathbb E_{m} \left\{{\sum_{k=1}^{K}}\left[ {\nabla _{\vect\theta} }{\vect\pi _{\vect\theta} }\left( \vect s_k \right) {{\nabla _{\vect\theta} }{\vect\pi _{\vect\theta} }{{\left( \vect s_k \right)}^{\top}}{\vect w}} \right]\right\},
\end{align}
and the proposed MPC-LSTD-based DPG method is summarized in Algorithm \ref{algo}.
\IncMargin{0.5em}
\begin{algorithm}
\algsetup{linenosize=\small}
\small
\KwIn{vessel model, objective function, initial parameters $\vect\theta_0$}
\KwOut{locally optimal policy $\vect\pi_{{\vect\theta}^\star}$}
\Repeat
{\text{convergence}}
{\For{each episode \rm{in} $m$ \rm{episodes}}{initialize $\vect\eta_0$, $\vect\nu_0$\;
\While{${{{\left\| {\vect\eta_k  - {\vect\eta _d}} \right\|}^2_2} \leq d_{\mathrm{error}}}$}{
solve the MPC \eqref{eq:mpc} and get ${\vect y}^\star$\;
calculate and record the RL stage cost $L{\left(\vect s_k, \vect a_k\right)}$ according to \eqref{eq:rl cost} and the sensitivity ${\nabla_{\vect\theta}}{\pi_{\vect\theta} }{\left(\vect s_k\right)}$ according to \eqref{eq:sensetivity}\;
}}
calculate $\vect v$ according to \eqref{eq:lstdq_1}\;
calculate $\vect w$ according to \eqref{eq:lstdq_2}\;
update $\vect \theta$ according to \eqref{eq:new_theta}\;
}
\caption{\small MPC-LSTD-based DPG method}
\label{algo}
\end{algorithm}
\DecMargin{0.5em}
\begin{table}[htbp!]
\renewcommand\arraystretch{1.25}
\caption{\label{tab:table1} Parameters values.}
\centering
\begin{tabular}{cc||cc}
\hline
Symbol & Value & Symbol & Value\\
\hline
$\gamma,N,dt$ & $1,60,0.5$ &
$\tau_a,{\alpha}_{\mathrm{max}}$ & $\mathcal{N}(0,1$e$-3),\frac{17\pi}{18}$\\
${f_1}_{\mathrm{min,max}}$  & $-100,100$      &
${f_{2,3}}_{\mathrm{min,max}}$  & $0,200$      \\
$\rho,\varepsilon,\delta$ & $1,0.001,0.001$&
$\vect W$ & $\mathrm{diag}([1,1,1])$\\
$\vect \omega, \vect \omega_f$ & $[1,5,5]^\top$&
$c_{1,2,3}$ & $5,8,8$\\
$d,d_s,d_{\mathrm{error}}$ & $42.5,1,0.5$&
$N_o, m$ &$3, 10$\\
$r_0,r_1,r_2,r_3$ & $1,1.4,1.7,1.9$&
$\vect \eta_d$ & $[21.3,23.3,8.4]^\top$\\
${\vect \eta}_0$ & $[0,0,\frac{\pi}{4}]^\top$&
${\vect \nu}_0$ & $[0.4,0,0]^\top$\\
\hline
\end{tabular}
\end{table}
\begin{figure}[htbp!]
\centering
\includegraphics[width=0.35\textwidth]{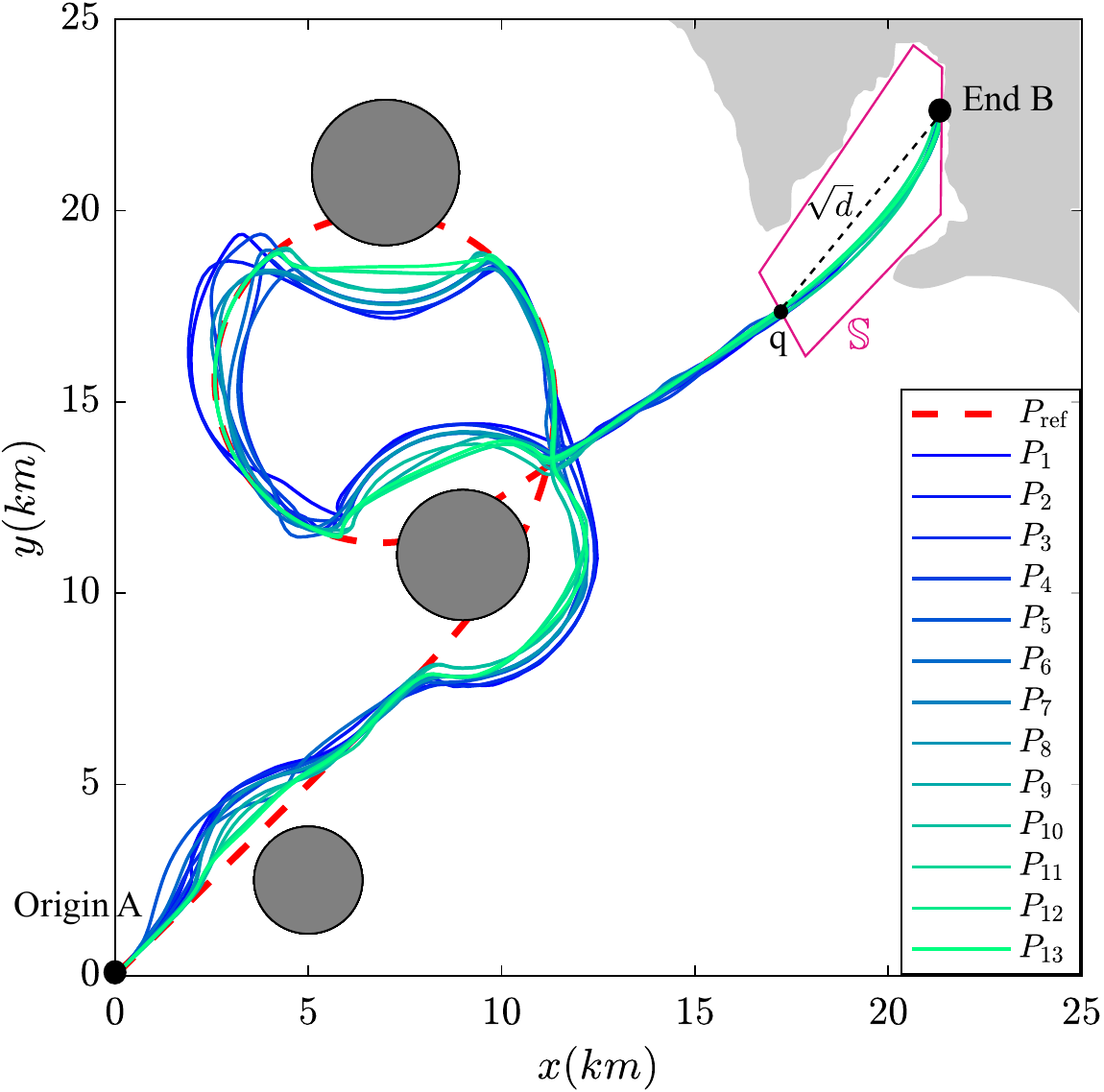}
\caption{Freight shipping paths from A to B. $P_{\mathrm{ref}}$: the reference path. $P_{1}$-$P_{13}$: the renewed path after each learning step.}
\label{fig:path}
\end{figure}
\begin{figure}[htbp!]
\centering
\includegraphics[width=0.4\textwidth]{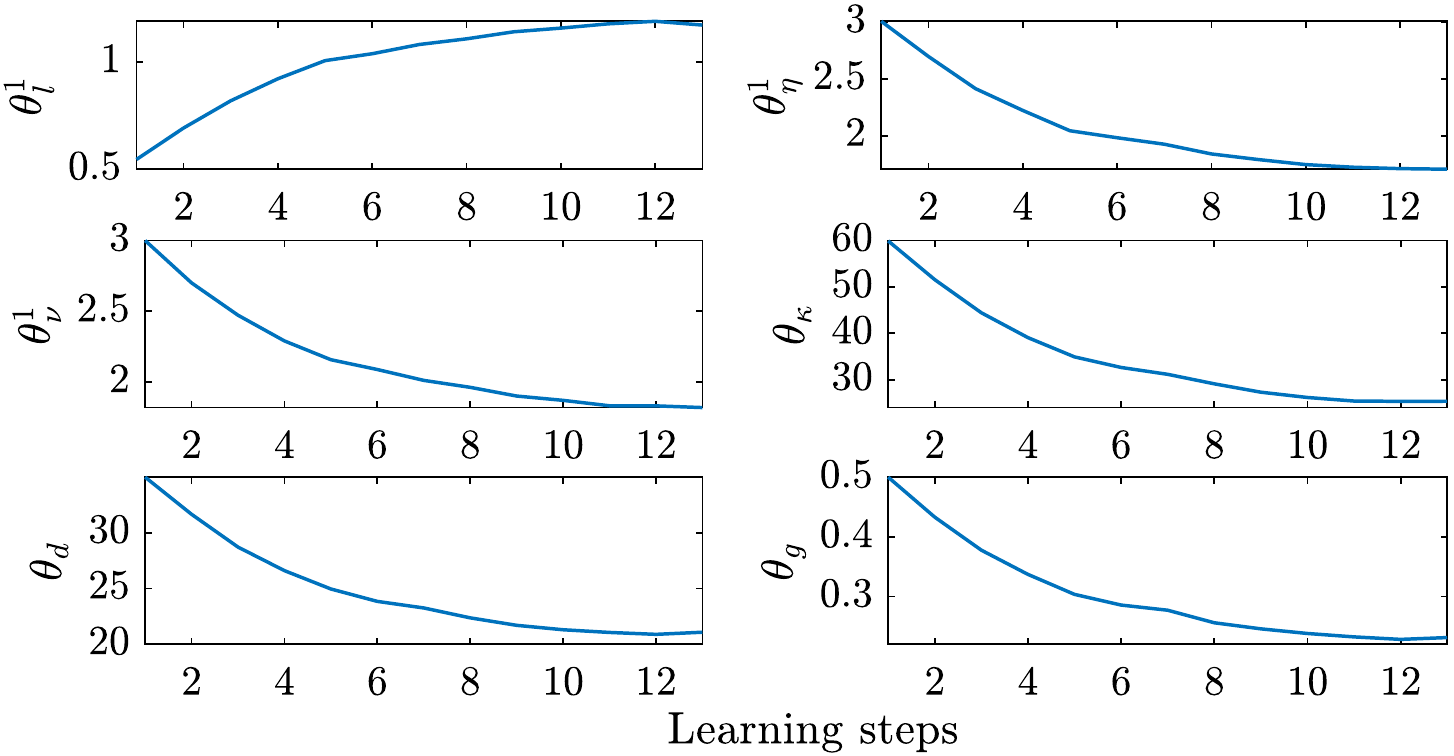}
\caption{Variations of some selected MPC parameters $\{\theta_l^1,\theta_{\eta}^1,\theta_{\nu}^1,\theta_\kappa,\theta_d,\theta_g\}$ over learning steps.}
\label{fig:theta}
\end{figure}
\begin{figure}[htbp!]
\centering
\includegraphics[width=0.398\textwidth]{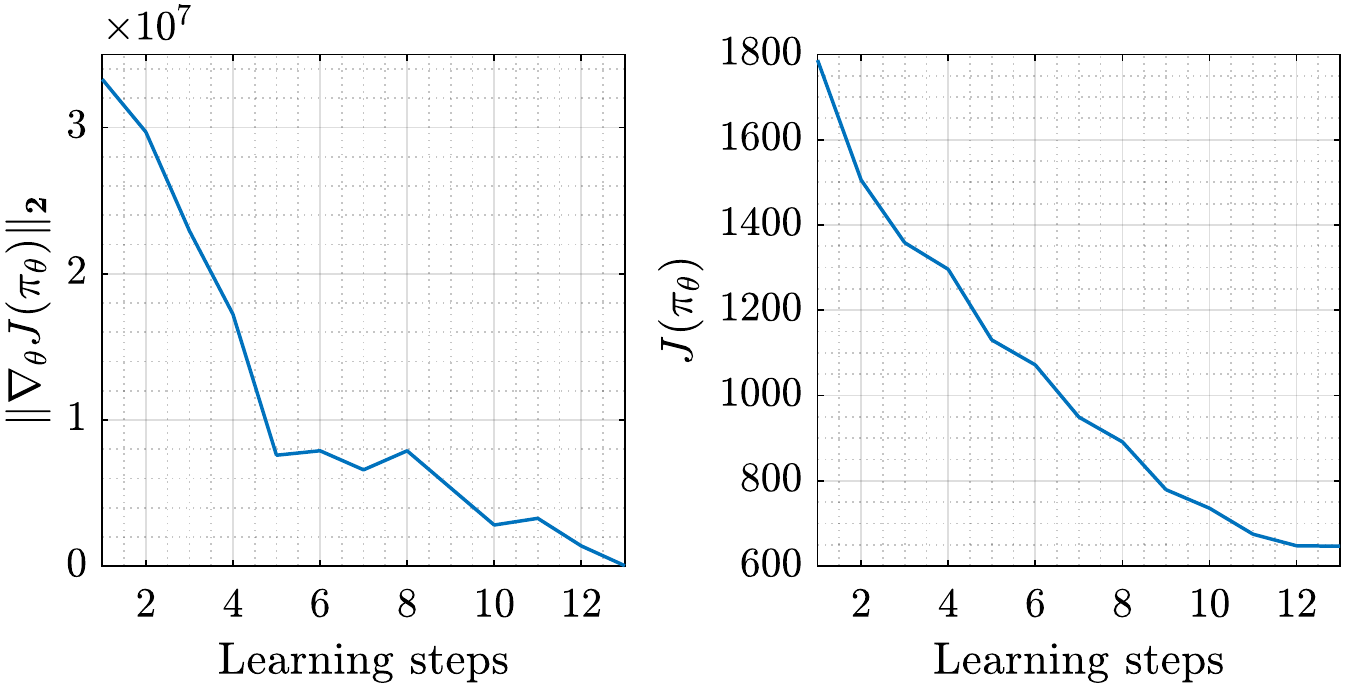}
\caption{Variations of the normed policy gradient $\|\nabla _{\vect\theta}J(\vect\pi _{\vect\theta})\|_2$ and the closed-loop performance $J(\vect\pi _{\vect\theta})$ over learning steps.}
\label{fig:performance}
\end{figure}
\begin{figure}[htbp!]
\centering
\includegraphics[width=0.368\textwidth]{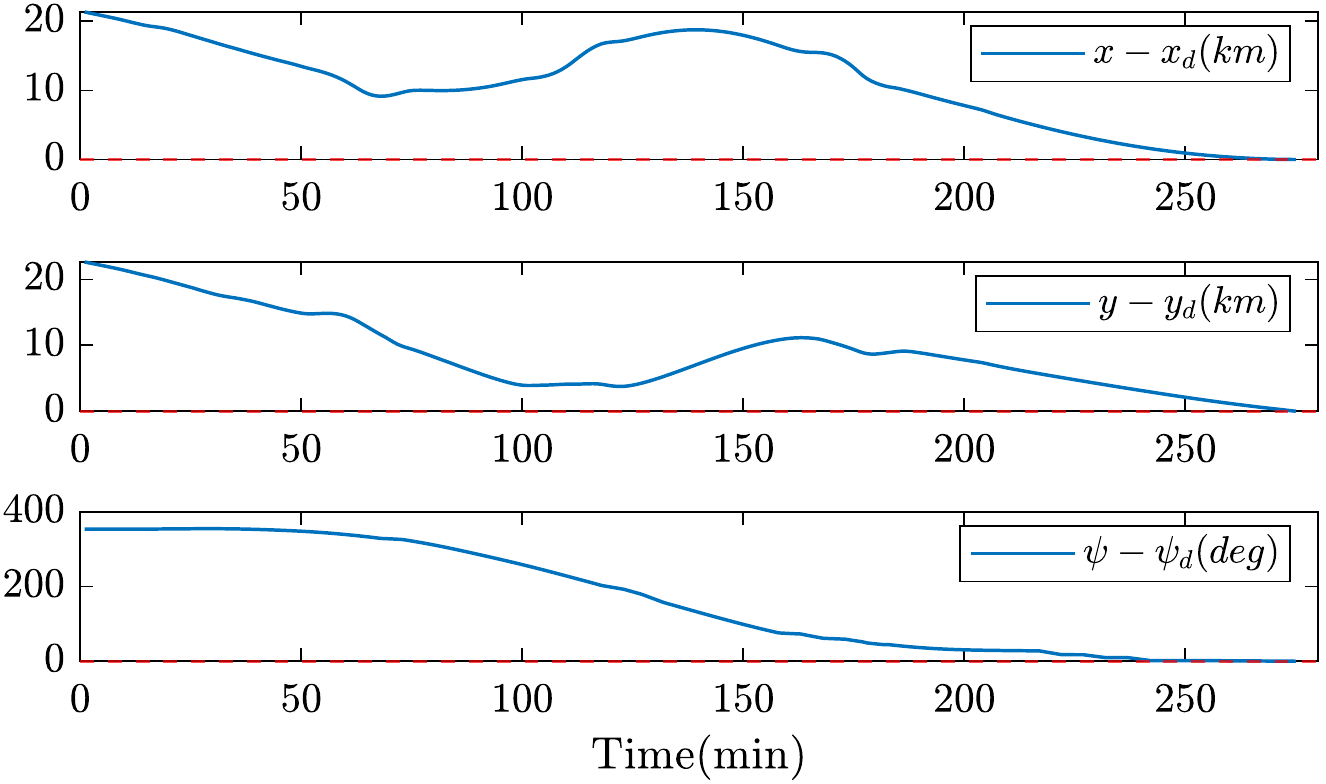}
\caption{Variations of the error $\vect\eta-\vect\eta_d$ with time under the learned policy $\vect\pi_{{\vect\theta}^\star}$. Red line: the desired value.}
\label{fig:pose}
\end{figure}
\begin{figure}[htbp!]
\centering
\includegraphics[width=0.38\textwidth]{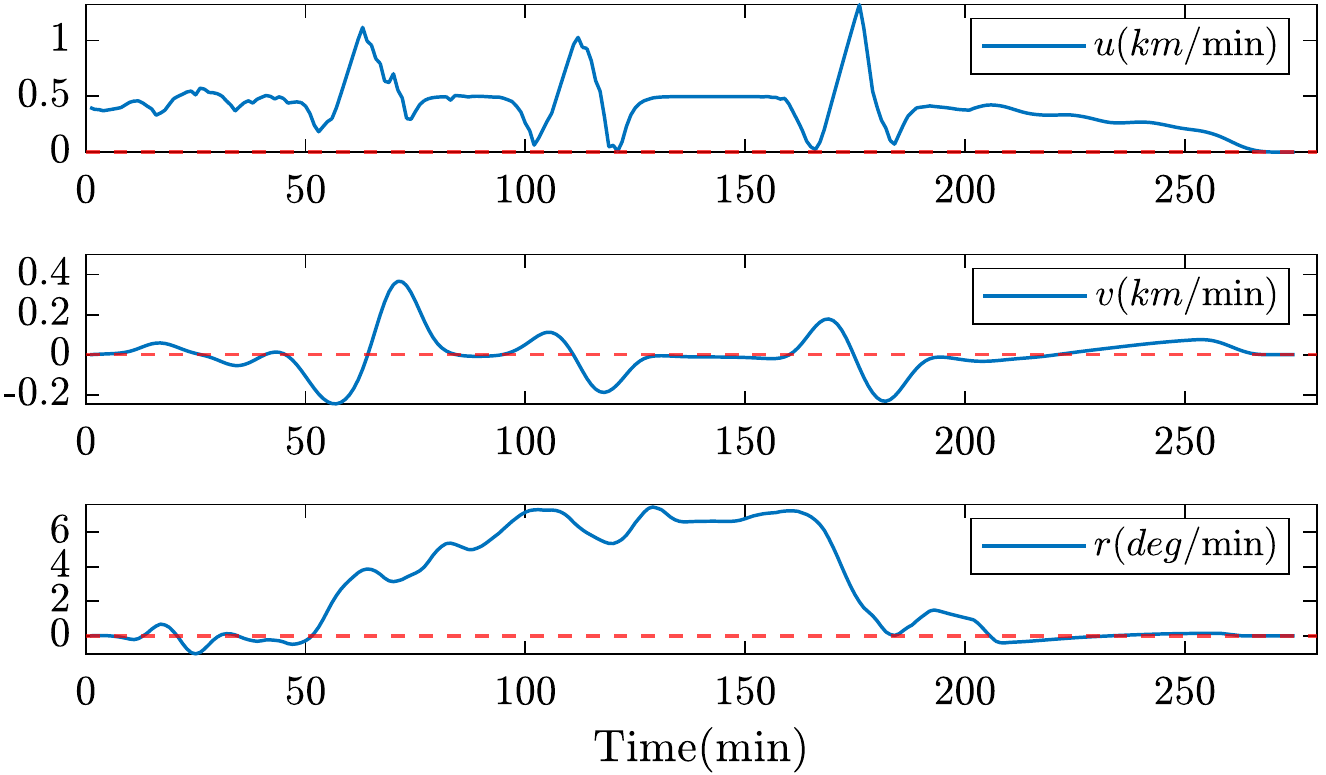}
\caption{Variations of the vessel velocity $\vect\nu$ with time under the learned policy $\vect\pi_{{\vect\theta}^\star}$. Red line: the desired value.}
\label{fig:velocity}
\end{figure}
\begin{figure}[htbp!]
\centering
\includegraphics[width=0.4\textwidth]{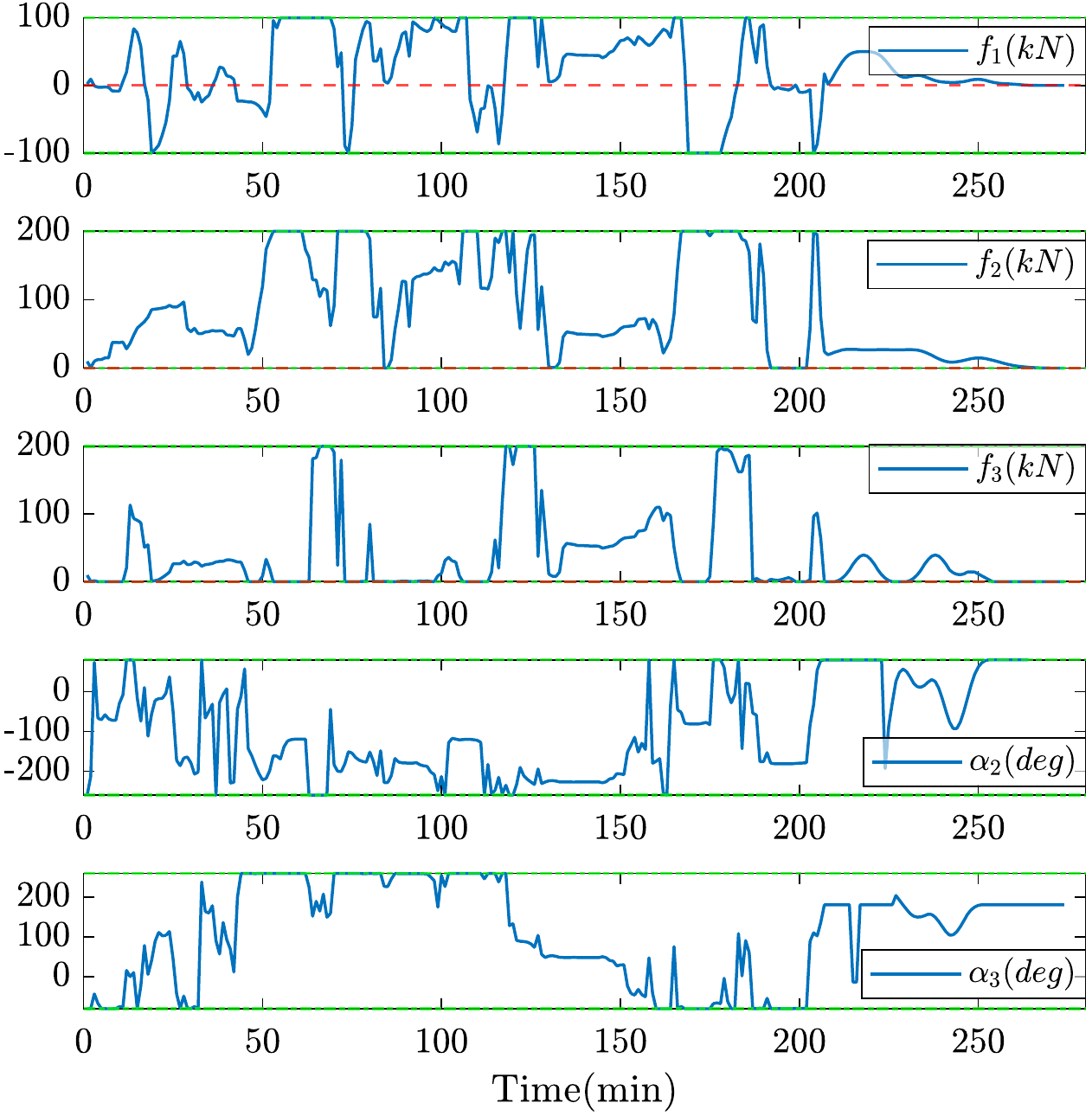}
\caption{Variations of the thruster force $\vect f$ and thruster angle $\vect \alpha$ with time under the learned policy $\vect\pi_{{\vect\theta}^\star}$. Green line: the constraint value.}
\label{fig:input}
\end{figure}

\section{Simulation}
\label{sec:simu}
In this section, we show the simulation results of an ASV freight mission problem using the introduced MPC-based RL method. We choose the initial parameters vector as ${\vect\theta}_0={\{\vect{0.55},\vect{3},\vect{3},\vect{1e{-7}},60,35,0.5\}}$, where the bold numbers represent constant vectors with suitable dimension. Other parameters values used in the simulation are given in Table \ref{tab:table1}. 

\par Figure \ref{fig:path} shows the prescribed reference path and the thirteen shipping paths updated after each learning step. The last path $P_{13}$ is obtained under the final learned policy $\vect\pi_{{\vect\theta}^\star}$ with an episode length of $K=550$. It is worth noting that, although we say that if the parametrization is rich enough, the MPC scheme can generate the optimal policy, this is a theoretical result. In practice, the assumption of a ``rich enough" parametrization is typically not satisfied. Other practical issues can come in the way of optimality such as, e.g., the local convergence of the RL algorithm and of the solver treating the MPC scheme. Addressing these potential issues typically requires good initial guesses. Although these are often available in the MPC context, we can only claim that the final learned policy $\vect\pi_{{\vect\theta}^\star}$ obtained from the converged parameters $\vect \theta^\star$ is locally optimal. This observation applies to most RL techniques. Following the reference path $P_{\rm{ref}}$ defined from the origin $A$ to the point $q$, the vessel departs from $A$ and passes through three obstacles to reach $q$. At the point $q$, where ${{\left\| {\vect\eta_k - {\vect\eta _d}} \right\|}^2_2} = d$, the vessel transits from path following to docking. The vessel eventually stops at the end $B$ with zero velocities and thruster forces, and has no collision with the quay (within the safety operation region $\mathbb{S}$) during the docking process. It can be seen that in the first few paths ($P_1$-$P_4$), the ship does not follow $P_{\rm{ref}}$ precisely, and is relatively far away from the three obstacles when it bypassed them. After learning, such as in the $P_{13}$, the ship follows closely the reference route, and the distance when avoiding obstacles is also reduced.

\par Figure \ref{fig:theta} shows the convergences of the MPC parameters $\vect\theta$ over learning steps ($\vect \theta^\star$ represents the converged parameters). Note that $\theta_l^1$ is the first element of $\vect\theta_l$, and the same fashion for others. It can be seen that the initial value of $\vect\theta_l$ is relatively small, and the initial values of $\vect\theta_{\eta},\vect\theta_{\nu},\theta_\kappa, \theta_d$ are relatively large. Therefore, in the MPC cost \eqref{eq:mpc_cost}, the terminal cost weights more than the stage cost, i.e., docking is regarded as more important than path following. Consequently, the path following performance is relatively poor in the initial episodes, and then gets improved as $\vect\theta_l$ increases and $\vect\theta_{\eta},\vect\theta_{\nu}, \theta_\kappa, \theta_d$ decrease. In addition, the initial value of $\theta_g$ is large, which means that the ship must be very far away from the obstacles. However, this is unnecessary under the premise of ensuring the safe distance $d_s$. To reduce the cost, RL gradually reduces $\theta_g$, and therefore results in what we have in Fig. \ref{fig:path}: the distance for avoiding obstacles tends to decrease over learning. 

\par The variations of the normed policy gradient $\|\nabla _{\vect\theta}J(\vect\pi _{\vect\theta})\|_2$ and the closed-loop performance $J(\vect\pi _{\vect\theta})$ are displayed in Fig. \ref{fig:performance}. As can be seen, the policy gradient converges to near zero and the performance is improved significantly over learning. Figure \ref{fig:pose} illustrates the variations of error between the vessel pose state $\vect\eta$ and the desired docking state $\vect\eta_d$ under the learned policy $\vect\pi_{{\vect\theta}^\star}$. Figure \ref{fig:velocity} presents the variations of the vessel velocity $\vect\nu$ with time under the policy $\vect\pi_{{\vect\theta}^\star}$. The red dash lines in these two figures represent the zero-valued reference lines. It can be seen that both the pose error and velocity converge to the red dash lines, which signifies a satisfactory docking. The variations of the vessel's thruster force $\vect f$ and thruster angle $\vect\alpha$ under policy $\vect\pi_{{\vect\theta}^\star}$ are exhibited in Fig. \ref{fig:input}. The green lines stand for the constraint values. As can be seen, both the forces and the angles obey their constraints, and when approaching the endpoint, the forces decline to zero and the angles remain constant. 
\section{Conclusion}
\label{sec:conc}
This paper presents an MPC-based RL method for the ASV to accomplish a freight mission, which includes collision-free path following, autonomous docking, and an ingenious transition between them. We use a parameterized MPC-scheme as the policy approximation function, and adopt the LSTD-based DPG method to update the parameters such that the closed-loop performance gets improved with learning. For future works, we will further validate our proposed method by realizing the experimental implementations.
\bibliographystyle{IEEEtran}
\bibliography{reference}
\end{document}